\documentclass[english]{article}
\usepackage[T1]{fontenc}
\usepackage[latin9]{inputenc}
\usepackage[a4paper]{geometry}
\usepackage{units}
\usepackage{textcomp}
\usepackage{amsmath}
\usepackage{graphicx}
\usepackage{esint}

\makeatletter
\newcommand{\lyxaddress}[1]{
\par {\raggedright #1
\vspace{1.4em}
\noindent\par}
}

\makeatother

\usepackage{babel}
\begin{document}

\title{The connection between Superconductors of Type I and II due polarons
and Itinerant Antiferromagnetism in weak's coupling regime.}

\author{Paulo S. A. Bonfim}

\maketitle

\lyxaddress{Universidade Unigranrio, Rua Prof. José de Souza Herdy, 1.160, 25071-202,
Duque de caxias, RJ ,Brazil.}
\begin{abstract}
In the weak regime, we suggest a model where superconductivity types
I and II bind assuming an adiabatic hypothesis where the frequency
of oscillation of free electrons is much smaller than of polarons
coupled to phonons, and an itinerant antiferromagnetism appears. At
the end of the article we derive an state's equation which is tested
according to generic experimental data. 
\end{abstract}
The discovery of new classes of superconductors has been a constant
in recent years{[}1{]}. However, the type I and type II superconductors
keep the same basic characteristic of its description in BCS model:
almost perfect diamagnetic behavior. Under controlled conditions ,
keeping the temperature mostly below the transition, worth the London
equations that are accurate for Type I superconductors but not realize
some properties of type II superconductors, especially the penetration
of the magnetic field in the high transition temperature of some (especially
of cupric oxides and recently the Iron arsenides) {[}2{]}. Therefore,
it is desirable that any theory that allows for the critical temperature
of superconductors of type II, in some limit, reduces to the usual
BCS state\textasciiacute{}s equation. There have been developments
in this direction for a perturbation theory in weak coupling, Eliashberg
et al {[}3{]} , but was abandoned due to the idea that superconductivity
of type II have separate origin of type I. Series of recent experiments,
especially EPR {[}4{]}, try to prove set of theories that assume strong
coupling and strongly localized nature where the most used techniques
consist of Green's functions in appropriate bases (For example helicity)
whose solutions are exact {[}5{]} by deviating from the perturbative
techniques. These approaches are derived from the great success of
quantum field theory when applied to low-dimensional systems and in
particular Strong Couplings for Heavy Fermions called where there
is also large local magnetic influences. The Hubbard Hamiltonian applies
very well to this kind of problems especially for Cold Atoms and Phases
ferromagnetic and antiferromagnetic.

But the same experiments in EPR {[}4{]} in compounds of type YBCO
also remind us of features which would fit a perturbative approach:
possibility of itinerant antiferromagnetism, coexistence of holes,
electron and polarons on the same phenomenon. More recently, we observed
a type II superconductor where the spatial symmetry was not broken,
there was no privileged plane for the superconducting conduction {[}2{]}.

In this paper we propose a simple model as another of the many mechanisms
that compose the type superconductivity II, where the wave S of Cooper's
pairs receives boost with tunneling Josephson of free electrons in
barriers produced by polarons phenomenologically{[}6{]}. Such barriers
produce a state of Itinerant Antiferromagnetism in a dynamic electrons\texttimes{}holes.
In the case of cupric oxide, holes are produced by oxygen which reduces
the copper. Once reduced ion Cu\textsuperscript{+2} suffers an effect:
the proximity of the orbitals 3p\textsuperscript{6} and 3d\textsuperscript{9}possibly
produces hybridization. The íon Cu\textsuperscript{+2}have one of
the greatest potential third ionization of the known elements. This
hybridization would increase the overall energy of the orbital p and
d away from the atom's nucleus, approaching the orbitals of the ions
Cu\textsuperscript{+2}neighbors, but with a hole produced by oxygen.
And producing the following effects: The inner orbital hybridized
form a potential barrier to the free electrons since it would be quite
filled, the hole produced simultaneously by oxygen would form a well
and a few free electrons would be released. This structure would vibrate
with frequency close to that of Phonons network and be coupled to
the same, very close to a polaron. This state would be frozen (adiabatic
hypothesis) relative to the free electrons because the oscillation
frequency of the polaron would be much smaller than the free electrons
($\omega_{P}\ll\omega_{e}$).

Selective doping elements donors increase the number of barriers to
a certain limit where excess barriers and reducing electrons available
diminish the effect of Josephson tunneling in the state IAF (Itinerant
AntiFerromagnetism).Therefore, the polaron electron-hole generates
new 'gap' that can be additive to 'gap' BCS, sometimes competing and
other reinforcing and increasing its critical temperature.

\section{Frozen Phonon x Polaron State - Adiabatic Hipotesis. }

How we work in perturbation hypothesis we keep the original BCS Hamiltonian
and assuming a structure of barriers is generated by the dynamics
electrons\texttimes{}hole where the coupling phonon\texttimes{}polaron
produces a structure barriers. The validity of this hypothesis derives
from the fact that the oscillation frequency of polarons be much smaller
than the free electrons involved: $\omega_{P}\ll\omega_{e}$. Soon,
though dynamic, the proposed structure is frozen in relation to frequency
of free electrons, ie, our model arises from an adiabatic hypothesis.
Follows the figure:

\includegraphics[width=14cm,height=7cm]{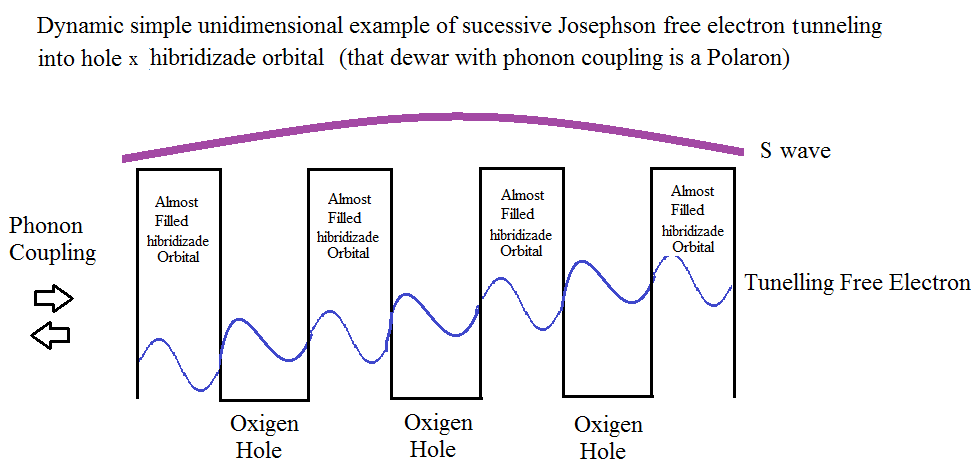} 

The latest models for cupric oxides adopt the strategy of making the
odd dispersion of free electrons of the crystal as a function in momentum
space{[}7{]}, we will adopt the same strategy: $\epsilon_{-k}=-\epsilon_{k}$
. This consideration derives from the fact that the wave function
be symmetric BCS for long-range interactions. But locally, the short
range of each individual electron wave must be antisymmetric whereas
the order parameter that generates the 'gap superconductor: $\left\langle c_{\uparrow k}^{\dagger}c_{\downarrow-k}^{\dagger}\right\rangle =\left\langle c_{k}^{\dagger}c_{-k}^{\dagger}\right\rangle \neq0$
, and invariably, $\left\langle c_{\downarrow k}^{\dagger}c_{\uparrow-k}^{\dagger}\right\rangle =0$
. Following similar reasoning any dispersion relation derived from
global interactions, long-range, will be considered even function
in momentum space: $F_{-k}=F_{k}$.

\section{Hamiltonian}

With great success Scrieffer \& Wolff {[}8{]} used a canonical transformation
to include second order effects of an exchange. Kondo {[}9{]} e Appenbaum
{[}10{]} ,treated perturbative form of the scattering of electrons
by magnetic impurities.The dynamic nature of superconductivity does
not seem limited by the low dimensionality of the system, then consider
that the superconducting phase is not limited sites as suggested by
models developed in recent years {[}11{]}. In particular, we will
not use a local model sites (Hubbard) and define a Hamiltonian in
momentum space.

\begin{alignat*}{1}
H=\sum_{k}\epsilon_{k}c_{k}^{\dagger}c_{k}+\varepsilon_{T}c_{T}^{\dagger}c_{T}+U\sum_{k}c_{k}^{\dagger}c_{k}c_{-k}^{\dagger}c_{-k} & +
\end{alignat*}

\begin{equation}
+\sum_{k}\left[V_{k,T}c_{k}^{\dagger}c_{T}+V_{k,T}^{*}c_{T}^{\dagger}c_{K}\right]
\end{equation}

In this Hamiltonian in particular see how the bound state produced
by electron tunneling $\epsilon_{T}c_{T}^{\dagger}c_{T}$ the structure
of the frozen barrier polaron (figure 1) affects the BCS interaction
and how the mixing of populations via 'exchange' affects the critical
temperature of the superconducting state. The terms mixture $V_{k,T}c_{k}^{\dagger}c_{T}$
are the first order $V\sim V_{k,T}\sim V_{k,T}^{*}$ we consider very
near and we will make canonical transformation in the second order
{[}8{]} whereas that $V\ll\varepsilon_{k}+U$ define convergence's
range $r=\nicefrac{\Gamma}{\left(\varepsilon_{k}+U\right)}$ where
$\Gamma=\pi N(\varepsilon_{T})AVE$ , see how the set of quasi-bound
states via tunneling affects the BCS term $U\sum_{k}c_{k}^{\dagger}c_{k}c_{-k}^{\dagger}c_{-k}$,
just only interested in the $V^{2}$ terms.

\section{The Canonical transformation}

We use the Baker-Hausdorff lemma so that only the terms proportional
to $V^{2}$ 'mix' the terms of sector $U$ of the Hamiltonian. We
will eliminate the terms proportional to $V$:

\begin{equation}
\overline{H}=H_{0}+\nicefrac{1}{2}\left[\hat{S},V\right]+\nicefrac{1}{3!}\left[\hat{S},\left[\hat{S},V\right]\right]...
\end{equation}

We have a simple way for the operator$\hat{S}$ {[}8{]} :

\begin{equation}
\hat{S}=\sum_{k'}\frac{V_{k'}n_{-k'}c_{T}^{\dagger}c_{k'}}{\epsilon_{T}-(\epsilon_{k}+U)}
\end{equation}

Considering $n_{m}=c_{m}^{\dagger}c_{m}$ and the usual relations
$\left\{ c_{m}^{\dagger},c_{l}\right\} =\delta_{ml}$; $\left\{ c_{m}^{\dagger}c_{l}^{\dagger}\right\} =\left\{ c_{m}c_{l}\right\} =0$.
We will use only the term $\overline{H}=H_{0}+\nicefrac{1}{2}\left[\hat{S},V\right]$
the perturbative series, easily found;

\begin{alignat*}{1}
\overline{H}=\left[\epsilon_{T}+\frac{1}{2}\sum_{k}\frac{V_{k}^{2}n_{-k}}{\epsilon_{T}-(\epsilon_{k}+U)}\right]n_{T} & +
\end{alignat*}

\begin{equation}
+\sum_{k}\epsilon_{k}n_{k}+\left[U-\frac{1}{2}\sum_{k}\frac{V_{k}^{2}}{\epsilon_{T}-(\epsilon{}_{k}+U)}\right]n_{k}n_{-k}
\end{equation}

In the case, $n_{T}\ll N_{k}=\sum_{k}n_{k}$ incorporated $E_{T}+\frac{1}{2}\sum_{k}\frac{V_{k}^{2}n_{-k}}{\epsilon_{T}-(\epsilon{}_{k}+U)}\sim E_{T}$
in $\sum_{k}\epsilon_{k}n_{k}$ . We got to the effective Hamiltonian,
which reversed $\epsilon_{T}$ with $\epsilon_{k}$.

\begin{equation}
H_{EFF}=\sum_{k}\left[\epsilon_{k}n_{k}+\left(U+\frac{1}{2}\sum_{k}\frac{V_{k}^{2}}{\epsilon_{k}+(U-\epsilon_{T})}\right)n_{k}n_{-k}\right]
\end{equation}

According to our previous considerations on two distinct populations:
influence of fermionic about 'odd' dispersion $\epsilon_{k}$ 'local,
short-range' and Bosonic (Copper's pairs) about the effects of Coulomb
and tunneling of the quasi-bound states of the potential of the polaron
binding.

\section{Successive tunneling and the order parameter particle \texttimes{}
hole.}

We define an order parameter $\triangle=\left\langle c_{k}^{\dagger}c_{-k}\right\rangle $
,   Hole \texttimes{} particle, consistent with our model, we will
introduce later phonons in the 'gap' equation with Debye energy's
cut-offs.

We can consider that$\triangle\neq0.$ We rewrite (5) in terms of
this parameter. Plus, as we consider the energy $\epsilon_{T}\ll U$
. Locally, the denominator of the term $\frac{1}{2}\sum_{k}\frac{V_{k}^{2}}{\epsilon_{k}+(U-\epsilon_{T})}$
slightly affects the interaction BCS $U$ . Thus we can group the
term $U+\frac{1}{2}\sum_{k}\frac{V_{k}^{2}}{\epsilon_{k}+(U-\epsilon_{T})}=\triangle_{NL},$
(return later to term), almost constant for times very close around
the time of Fermi  $k\sim k_{F}$   using the BCS weak coupling hypothesis.
Manipulating (5) and considering the order parameter:

\begin{alignat*}{1}
H_{EFF}=\sum_{k}\left[\epsilon_{k}c_{k}^{\dagger}c_{k}+\triangle_{NL}c_{k}^{\dagger}c_{k}(\hat{1}-c_{-k}c_{-k}^{\dagger})\right] & =
\end{alignat*}

\begin{equation}
=\sum_{k}\left[(\triangle_{NL}+\epsilon_{k})c_{k}^{\dagger}c_{k}+\triangle_{NL}c_{k}^{\dagger}c_{-k}c_{-k}^{\dagger}c_{k}\right]
\end{equation}

In our model we adopt the order parameter electron\texttimes{}hole$\left\langle c_{k}^{\dagger}c_{-k}\right\rangle \neq0$,
 of ( 6 ) we find:

\begin{equation}
H_{EFF}=\sum_{k}\left[(\triangle_{NL}+\epsilon{}_{k})c_{k}^{\dagger}c_{k}+\triangle_{NL}\triangle c_{-k}^{\dagger}c_{k}\right]
\end{equation}

Now we use the assumption of the 'quasi bound states' QBS to manipulate
summations and use the parity of the dispersions around $k_{Fermi}$,
We impose $k_{Fermi}=0$:

\begin{equation}
\sum_{k}=\sum_{-k}+\sum_{+k}
\end{equation}

The first summation on the right is opposite to the second sum to
the right times. This is possible only with the hypothesis of QBS
state, which transforms the equation ( 7 ) , as follows:

\begin{alignat*}{1}
H_{EFF}=\sum_{k}\left[(\triangle_{NL}+\epsilon{}_{k})c_{k}^{\dagger}c_{k}+\triangle_{NL}\triangle c_{-k}^{\dagger}c_{k}\right] & +
\end{alignat*}

\begin{equation}
+\sum_{-k}\left[(\triangle_{NL}+\epsilon{}_{k})c_{k}^{\dagger}c_{k}+\triangle_{NL}\triangle c_{-k}^{\dagger}c_{k}\right]
\end{equation}

To use our approach parity of dispersions local and non-local we unified
de summation (8) around the Fermi energy, Taking the conversion $k\rightarrow-k$
in the sector $\sum_{-k}$ of equation (9) simultaneously embracing
$\triangle\sim\triangle^{*}$, found :

\begin{alignat*}{1}
H_{EFF}=\sum_{k}\left[(\triangle_{NL}+\epsilon{}_{k})c_{k}^{\dagger}c_{k}+\triangle_{NL}\triangle c_{-k}^{\dagger}c_{k}\right] & +
\end{alignat*}

\begin{alignat*}{1}
+\sum_{k}\left[(\triangle_{NL}+\epsilon{}_{-k})c_{-k}^{\dagger}c_{-k}+\triangle_{NL}\triangle^{*}c_{k}^{\dagger}c_{-k}\right] & =
\end{alignat*}

\begin{alignat*}{1}
=\sum_{0}^{k}[(\triangle_{NL}+\epsilon{}_{k})c_{k}^{\dagger}c_{k}+\triangle_{NL}\triangle c_{-k}^{\dagger}c_{k} & +
\end{alignat*}

\begin{alignat*}{1}
+ & (\triangle_{NL}+\epsilon{}_{-k})c_{-k}^{\dagger}c_{-k}+\triangle_{NL}\triangle^{*}c_{k}^{\dagger}c_{-k}]
\end{alignat*}

Around the Fermi energy, assuming $k_{Fermi}=0$, with $\epsilon_{-k}=-\epsilon_{k}$
for dispersions of free electrons (local influences) and $F_{-k}=F_{k}$
for Non local and global influences, we arrive at the same Hamiltonian
of ITINERANT ANTIFERROMAGNETISM {[}12{]}:

\begin{alignat*}{1}
H_{EFF}=\sum_{0}^{k}[(\triangle_{NL}+\epsilon{}_{k})c_{k}^{\dagger}c_{k}+(\triangle_{NL}-\epsilon{}_{k})c_{-k}^{\dagger}c_{-k} & +
\end{alignat*}

\begin{equation}
+\triangle_{NL}\triangle\left(c_{-k}^{\dagger}c_{k}+c_{k}^{\dagger}c_{-k}\right)]
\end{equation}

\section{Itinerant Antiferromagnetism, Bogoluibov Transform and Gap equation.}

The goal of our model is to link the particle x hole dynamics , with
phonon x polaron and itinerant antiferromagnetism. Let's perform a
transformation Bogoluibov and then retrieve $\triangle_{NL}$ with
$\theta=\theta_{k}$:

\begin{equation}
\left\{ \begin{array}{c}
c_{k}^{\dagger}=A_{k}^{\dagger}sen\theta-A_{-k}^{\dagger}cos\theta\\
c_{k}=A_{k}sen\theta-A_{-k}cos\theta\\
c_{-k}^{\dagger}=A_{-k}^{\dagger}sen\theta+A_{k}^{\dagger}cos\theta\\
c_{-k}=A_{-k}sen\theta+A_{k}cos\theta
\end{array}\right.
\end{equation}

The rotation allows us to diagonalize (10), defining new base: $u_{k}=sen\theta$
e $v_{k}=cos\theta$ , manipulating the sums as in (8) using the hypothesis
QBS:
\begin{equation}
H_{EFF}^{D}=\sum_{k}\left(\triangle_{NL}-\epsilon_{k}cos2\theta-\triangle_{NL}.\triangle sen2\theta\right)A_{k}^{\dagger}A_{k}
\end{equation}

Using the free energy HelmHoltz $F=\left\langle H_{EFF}^{D}\right\rangle -TS$,
taking its minimum at the new base $\frac{\partial F}{\partial\theta_{k}}=\frac{\partial\left\langle H_{EFF}^{D}\right\rangle }{\partial\theta_{k}}=2\epsilon_{k}sen2\theta_{k}-2\triangle sen2\theta_{k}$
, and subjecting (12) to a thermal bath $(1-2f_{k})$ being $f$$_{k}=\frac{1}{e^{\nicefrac{\epsilon}{kT}}+1}$,
we get the equation of 'Gap' to our system:

\begin{equation}
\epsilon_{k}tg2\theta_{k}=\triangle_{NL}.\triangle
\end{equation}

Fixing the free part and taking the thermal average $\triangle_{NL}=U+\frac{1}{2}\sum_{k}\frac{V_{k}^{2}}{\epsilon_{k}+(U-\epsilon_{T})}$:
setting usually via hypothesis BCS: $tg2\theta_{k}=\frac{\triangle_{k}}{\epsilon_{k}}$

\begin{equation}
\triangle_{k}=\sum_{k'}\left(U+\frac{1}{2}\frac{V^{2}}{\epsilon_{k'}+(U-\epsilon_{T})}\right).\triangle.(1-2f_{k'})
\end{equation}

In the above equation so that there is self-consistent solution, the
right fixes the left side that gets its share 'free' fixed according
to BCS hypothesis: $tg2\theta_{k}=\frac{\triangle_{k}}{\epsilon_{k}}$.
Here we assume, analogous to the BCS model, which $U$ and $V_{k}$
are smooth functions on the closed Fermi energy range. So we can solve
(14) an explicit form is required for the order parameter$\triangle$.
The physical sense of the parameter defines its shape as a function
of $k$, using as a reference text De Gennes {[}13{]} assume as a
solution to the Josephson tunneling: $E(k)=E_{0}+Jcosk\rightarrow E(\theta_{k})=E_{0}+Jcos2\theta_{k}$
in our model. Adapting our basic and simple form of a function $\theta_{k}=k$.
The wave packet in momentum space is obtained directly from$\frac{\partial E(\theta_{k})}{\partial\theta_{k}}=-2Jsen2\theta_{k}$,
ie, the antiferromagnetic order parameter has its origin bound states
formed by tunneling:

\begin{equation}
\triangle=\left\langle c_{k}^{\dagger}c_{-k}\right\rangle =\frac{\partial E(\theta_{k})}{\partial\theta_{k}}=-2Jsen2\theta_{k}
\end{equation}

Thus we can rewrite the equation of 'Gap':

\begin{equation}
\triangle_{k}=-2J.\sum_{k'}\left(U+\frac{1}{2}\frac{V^{2}}{\epsilon_{k'}+(U-\epsilon_{T})}\right).sen2\theta_{k}.(1-2f_{k'})
\end{equation}

But according to BCS hypothesis: $sen2\theta_{k}=\frac{\triangle_{k}}{\sqrt{\epsilon_{k}^{2}+\triangle_{k}^{2}}}$:

\begin{equation}
\triangle_{k}=-2J.\sum_{k'}\left(U+\frac{1}{2}\frac{V^{2}}{\epsilon_{k'}+(U-\epsilon_{T})}\right).\frac{\triangle_{k}}{\sqrt{\epsilon_{k'}^{2}+\triangle_{k}^{2}}}.(1-2f_{k'})
\end{equation}

Note that with $V\rightarrow0$, equation (17) returns to the 'Gap'
original BCS equation. At this point we use $(1-2f_{k'})\rightarrow tgh\left(\frac{\epsilon}{kT}\right)$.
In this model without the introduction of phonons theres no physical
sense. Originally BCS theory introduces phonons with a 'cut-off' in
the continuous limit in the energy space is the 'Debye temperature':
$\theta_{D}.$ We adopt the ansatz that the Debye temperature assumes$\sim4\times T_{c}$,
ie, the factor $tgh\left(\frac{\epsilon}{kT}\right)\rightarrow1$,
for $k\theta_{D}>kT$, as an asymptotic limit. Therefore, for $\epsilon_{T}\sim cte$,
and $\triangle_{k}\sim cte$, turning to energy space in continuous
limit, (17) is in the form:

\begin{equation}
1=-2JN\left[U\int_{kT_{c}}^{k\theta_{D}}\frac{d\epsilon}{\epsilon}+\frac{V^{2}}{2}\int_{kT_{c}}^{k\theta_{D}}\frac{d\epsilon}{\epsilon^{2}}\right].tgh\left(\frac{\epsilon}{kT_{c}}\right)
\end{equation}

In the asymptotic limit we propose, and $N$ being the density of
states:

\begin{equation}
1=-2JN\left[U\int_{kT_{c}}^{k\theta_{D}}\frac{d\epsilon}{\epsilon}+\frac{V^{2}}{2}\int_{kT_{c}}^{k\theta_{D}}\frac{d\epsilon}{\epsilon^{2}}\right]
\end{equation}

Using $J_{0}=2J$, we find:

\begin{equation}
1=J_{0}\left[NU.ln\left(\frac{T_{c}}{\theta_{D}}\right)+\frac{NV^{2}}{2k}\left(\frac{1}{\theta_{D}}-\frac{1}{T_{c}}\right)\right]
\end{equation}

Assuming $U<0.$

\section{Conclusions and future developments.}

The equation of state (20) in the limit $V\rightarrow0$ and with
$J\rightarrow1$ recovers the original BCS equation considering $U<0$
. We determined an equation of state which returns the type I superconductivity
when the second-order effects due to the polaron disappears. The effect
is very similar to the Jahn-Teller effect. Relationship is established
between the itinerant antiferromagnetism and superconductivity of
type II. 

To test the equation we calculate some experimental values {[}14{]}.
In particular the YBCuO family. The values we used were: $T_{c}\sim100\, K$
; Debye Temperature $\theta_{D}\sim410\, K$ (watching the most massive
and abundant ion Cu) ; $NU\sim-0.66$ ; $J_{0}\sim1.14$ ; $K_{B}=8.617\times10^{-5}eV/K$
with $\frac{1}{2K}\sim5802$ . And we find $NV^{2}\sim0,00123.$

In our little test, the ratio between the BCS interaction and second-order
effect due to tunneling of electrons through structure of polarons
is$\frac{U}{V^{2}}\sim550$ ,ie, an order of magnitude$10^{3}.$ 

For future developments, we can study $U(\omega)$ e $V(\omega)$
like unique spatial configurations functions of ions and polarons,
respectively.

\end{document}